\documentclass[conference]{IEEEtran}
\IEEEoverridecommandlockouts

\usepackage{cite}
\usepackage{amsmath,amssymb,amsfonts}
\usepackage{algorithmic}
\usepackage{graphicx}
\usepackage{lipsum}
\usepackage{textcomp}
\usepackage{xcolor}
\def\BibTeX{{\rm B\kern-.05em{\sc i\kern-.025em b}\kern-.08em
    T\kern-.1667em\lower.7ex\hbox{E}\kern-.125emX}}

\usepackage[T1]{fontenc}
\def\doi#1{\href{https://doi.org/\detokenize{#1}}{\url{https://doi.org/\detokenize{#1}}}}

\usepackage{times}
\usepackage{pifont}

\usepackage{ulem}
 
\usepackage[linkbordercolor={1 1 1}]{hyperref}

\usepackage{eucal}

\usepackage{colortbl} 
\usepackage{enumerate}

\hypersetup{
breaklinks=true,
  colorlinks = false, 
  urlbordercolor=white,
  linkbordercolor=white,
  citebordercolor=white,
  runbordercolor=white,
  citebordercolor=white,
  filebordercolor=white,
  linkcolor=blue 
  }

\newcommand*{\trim}[1]{%
  \trim@spaces@noexp{#1}%
}

\newcommand{\bluetext}[1]{\textcolor{black}{#1}}

\newcolumntype{P}[1]{>{\centering\arraybackslash}p{#1}}
\newcolumntype{R}[1]{>{\raggedleft\arraybackslash}p{#1}}                                                                                                                                               
\newcolumntype{L}[1]{>{\raggedright\arraybackslash}p{#1}}  
\usepackage{comment}

\newcommand{\cmark}{\ding{51}}
\newcommand{\xmark}{\ding{55}}

\begin{document}

\title{Mobile Mental Health Apps: \\Alternative Intervention or Intrusion?}

\author{\IEEEauthorblockN{Shalini Saini}
\IEEEauthorblockA{\textit{Department of Computer Science} \\
\textit{Texas A\&M University}\\College Station, TX, USA \\s.saini@tamu.edu}
\and
\IEEEauthorblockN{Dhiral Panjwani}
\IEEEauthorblockA{\textit{IT- Department of Medicine} \\
\textit{Univeristy of Alabama at Birmingham}\\Birmingham, AL, USA \\dpanjwani@uabmc.edu}
\and
\IEEEauthorblockN{Nitesh Saxena}
\IEEEauthorblockA{\textit{Department of Computer Science} \\
\textit{Texas A\&M University}\\College Station, TX, USA \\nsaxena@tamu.edu}
}

\maketitle

\begin{abstract}

Mental health is an extremely important subject, especially in these unprecedented times of the COVID-19 pandemic. Ubiquitous mobile phones can equip users to supplement psychiatric treatment and manage their mental health. Mobile Mental Health (MMH) apps emerge as an effective alternative to assist with a broad range of psychological disorders filling the much-needed patient-provider accessibility gap. However, it also raises significant concerns with sensitive information leakage. The absence of a transparent privacy policy and lack of user awareness may pose a significant threat to undermining the applicability of such tools. We conducted a multifold study of - 1) Privacy policies (Manually and with \textit{Polisis}, an automated framework to evaluate privacy policies); 2) App permissions; 3) Static Analysis for inherent security issues; 4) Dynamic Analysis for threat surface and vulnerabilities detection, and 5) Traffic Analysis.  
       
      Our results indicate that apps’ exploitable flaws, dangerous permissions, and insecure data handling pose a potential threat to the users’ privacy and security. The Dynamic analysis identified 145 vulnerabilities in 20 top-rated MMH apps where attackers and malicious apps can access sensitive information. 45\% of MMH apps use a unique identifier, \textit{Hardware Id}, which can link a unique id to a particular user and probe users’ mental health. Traffic analysis shows that sensitive mental health data can be leaked through insecure data transmission. MMH apps need better scrutiny and regulation for more widespread usage to meet the increasing need for mental health care without being intrusive to the already vulnerable population.

\end{abstract}

\pagestyle{plain}
\label{sec:intro}

\section{Introduction}

As per the Centers for Disease Control and Prevention (CDC), from mid-2020 to early 2021 (after starting COVID-19 ), there is an increase in anxiety or depressive disorder from 36.4\% to 41.5\%. The percentage of unmet mental health care needs is increased from 9.2\% to 11.7\% \cite{vahratian2021symptoms}. Depression, anxiety, and other mental disorders can cripple everyday functioning and even can claim lives as suicide is the tenth leading cause of death in the U.S. \cite{cdc_deaths2018}. Winkler et al. found that COVID-19 increased the prevalence of major depressive disorder and suicide risk three times and almost doubled the current anxiety disorders \cite{winkler2020increase}. However, a shortage of providers may encourage people to seek alternatives more often for immediate help \cite{thomas2009county}. Ubiquitous mobile devices can help to bridge the patient-provider gap and health divide for underserved and hard-to-reach populations to manage their mental health needs \cite{proudfoot2013future}. As per Wang et al., MMH apps have the potential to improve mental health, but most of the currently available apps lack clinical evidence to support their efficacy \cite{wang2018systematic}. Mainstream clinical practice incorporating MMH apps is also challenging as research is open to ensure that technological vulnerabilities do not compromise the privacy and safety of patients \cite{east2015mental, neary2018state,zhou2019barriers}. 

Mobile app developers may need standard guidelines to develop a secure MMH app with the help of the medical community on the usability \cite{chan2015towards}. Selecting a secure and reliable MMH app from a vast pool of available MMH apps may be daunting for the end-user. A well-defined privacy policy may be the starting point for app users in making decisions balancing information-sharing and preserving privacy. No explicit declaration on apps’ permissions, information is collected, and intended usage of collected information makes it difficult to grade the apps on security and privacy parameters. Unfortunately, unintended biases towards mental disorders increase a victim’s vulnerability to unwanted or unknown disclosure of his/her mental disorder. 

There is a lack of reproducible rigorous scientific testing to support rapidly developed MMH apps and quick launches in awaiting market. Currently, there is no national standard for evaluating the effectiveness of the available hundreds of mental health apps \cite{nih_mh}. It is essential to maintain transparency in privacy policies to build the needed trust for the broader usage of MMH apps. Our contribution is manifold through this study as follows:
\begin{itemize}
	\item Analyzing the availability, accessibility, and deficiencies of MMH apps’ privacy policies.
	\item Dynamic analysis to explore the threat surface, dangerous permissions, and injection vulnerabilities.
	\item Static analysis to study the apps’ code files exposing major exploitable security and privacy flaws.
	\item Network traffic analysis to study insecure data transactions between an app and any third party.
\end{itemize} 

Our results show that 95\% of 20 Highly-Popular (HP) apps provide a privacy policy compared to 65\% of 20 Less-Popular (LP apps). From 65\% of the analyzed HP apps, unique device information can tie users to a specific mental disorder. \textit{Polisis} analysis found that 9 out of 10 apps are collecting Device IDs and IP Addresses, which can pose a potential threat to users’ privacy and security. Dynamic analysis reveals that 17.67\% of all 20 HP apps permissions are categorized as dangerous by the Android framework. 11 out of 20 HP apps use one or more dangerous permissions. In static analysis, 45\% of 20 HP MMH apps show \textit{Hardware Id Usage}, a unique permanent device Id, and thus poses a privacy risk. 20\% apps show \textit{Insecure TLS/SSL Trust Manager}, which may cause insecure network traffic to leak sensitive data. 

To the best of our knowledge, there is no existing work with a comprehensive analysis for privacy and security concerns specific to MMH apps targeting vulnerable populations with a higher possibility of sharing personal and sensitive mental health information. 

The following section presents the background of the issue and the need for the study. Then, we discuss the related work done in the field, followed by our study design and goals. After that, we present the methodology and data collection, results, and challenges. We conclude the study findings with a discussion and future directions. 

\label{sec:background}
\section{Background}

Mental illnesses impact one in five United States (U.S.) adults with having anxiety disorders as the most common mental illness. Every year, around 40 million U.S. adults of age 18 and older are affected by anxiety disorders. Anxiety disorders affect 25.1\% of U.S. teenagers. Even though anxiety disorders are considered highly treatable, treatment covering only 36.9\% indicates a substantial gap \cite{adaa}. A study of depression data from 1990 to 2017 by Liu et al. found an overall increase in depression globally and staying as a global public health issue \cite{liu2020changes}. Studies have reported that in the U.S., nearly \$193 billion in lost earnings each year is attributed to mental health disorders \cite{khushalani2018systematic}. 

Overburdened and overwhelmed providers and patients are not able to afford the services because of limited insurance participation by providers and unclear or no coverage by insurance providers \cite{cummings2015rates,thomas2009county}. 

\subsection{MMH Apps as Mental Health Intervention}
MMH apps can assist huge underserved populations with a broad spectrum of mental disorders, from a simple reminder to take medicine on time to sending automatic signals outside regarding a predictable near-future crisis \cite{luxton2011mhealth}. The MMH app is convenient, low-cost, anonymous, free from human biases, and provides round-the-clock assistance to reach out to people in need. Interactive or game-based app therapies can invoke users’ interest, especially in a younger generation, and can help them to follow through \cite{rowntree2019smartphone}. MMH app can be a great supporting tool to conventional in-person therapy sessions by providing supplement data \cite{nih_mh}.  

A large pool of MMH apps is available through iOS and Android app stores. If app stores can identify the scientific research-based MMH apps for users, It may add confidence to use the app \cite{giota2014mental}. User ratings are helpful but not necessarily valid for clinical usefulness backed up by research. Online resources like \textit{PsyberGuide} may help provide objective and actionable information for publicly available MMH apps \cite{neary2018state}. 

The privacy and security concerns are potential factors to hinder the widespread acceptability, and use of MMH apps \cite{olff2015mobile}. A real-life patient’s perspective suggests that apps can be utilized in psychiatric treatment with flexible use of apps rather than relying on a single condition-specific MMH app \cite{chiauzzi2019mental}. 
 
\subsection{MMH Apps Security and Privacy Issues}
Users’ privacy and security is significant concern regarding MMH apps. As per O’Loughlin et al., 68\% apps received unacceptable transparency scores on data security, and privacy policies of studied mobile apps for depression \cite{o2019reviewing}. A lack of industry regulations and standards makes it very difficult to evaluate these apps on security, privacy, and effectiveness. Non-standard development practices may produce an insecure, easily exploitable app that malicious attackers can target to breach. An absence of a privacy policy in almost half of the studied apps shows a significant concern of not informing users about data collected, data security, and data sharing \cite{parker2019private}. 

Leaked information can harm users financially, mentally, physically, or emotionally, even life-threatening to already overwhelmed users who do not know how to handle any additional mental burden. Most of these policies use legal “boilerplate” without saying much about what the app does precisely and what is involved in data collection, sharing, or app permissions. The typical average user is not even familiar with legal or technical jargon and has little or no ability to understand to be agreed on the details given. 

In the absence of HIPAA protection, the app company may collect and share healthcare-related data for usage that the patient never imagined. Although the proliferation of mental health technologies like condition-specific smartphone apps and wearable sensors continues to increase, evidence for their clinical utility, efficacy, and safety is generally lacking. All MMH apps need proper identification if falling under compliance or outside the scope. At present, data brokers may end up indefinitely owning the patient’s data and using it for a variety of purposes like the generation of FICO Medication Scores, targeted advertisements, or more considerable profiling efforts \cite{armontrout2016mobile}. Based on the highly sensitive nature and broad scope of usage, MMH apps need more rigorous and continuous evaluation to ensure the privacy and security of these apps’ users.

\label{sec:related work}
\section{Related Work}
Mobile app security and privacy issue are well-acknowledged by healthcare, legal authorities, and researchers. A 2017 study in rural India shows that privacy policy complexity may be a barrier to informed decision making \cite{powell2018complexity}. Reardon et al. presented how mobile apps can gain access to protected data without user consent by using both covert and side channels through exploiting the Android permission model and posing a threat to users' privacy and security \cite{reardon201950}. \bluetext{Device ID is reported only for .2\% of tested apps (172/88113) by \cite{reardon201950}, but our work shows that 45\% of top-rated MMH apps use Hardware ID, which indicates a higher risk for the target population. We present all observed dangerous permissions for the tested apps and their possible misuse, while Reardon et al. analyzed only a few dangerous permissions uncovering covert and side channels.} Au et al. found a fundamental trade-off between the stability of the permission specification and enforcing least-privilege with fine-grain permissions, which may impact developers' choices of designing MMH apps and allowing unnecessary permissions bundled with the necessary ones \cite{au2012pscout}.

Giota et al. describe how using mental health apps can be risky compared to consulting the therapist daily, including loss of data and theft through insecure devices and communication channels \cite{giota2014mental}. Robillard et al. focus on privacy policies and terms of agreements covering 319 MMH apps for both Android and iOS platforms. Their work shows that only 18\% iOS and 4\% Android apps have privacy policies with collecting user information by 92\% of studied privacy policies \cite{robillard2019availability}. O'Loughlin et al. highlight the absence of privacy policy, consent, transparency, data sharing, and difficulty of readability for the general population \cite{o2019reviewing}. This work emphasizes the absence of privacy policy in nearly half of 116 studied apps and the absence of not covering what personal information is being collected \cite{o2019reviewing}. Parker et al. conducted an empirical study for 61 apps, identifying that malicious apps or attackers can exploit `dangerous' app permissions to access sensitive information \cite{parker2019private}. 

Significant recommendations are about the transparency of privacy policies on data sharing, allowing users to opt-out from data collection, improved user interface, clinical trials, rigorous evaluation, and integration with EHR \cite{torous2019towards}. Parker et al. studied the MMH apps' privacy policies aligned with the local government policy. Interestingly, the suggested solution is for users to pay more attention to selecting safe and efficient apps \cite{parker2019hot}. Papageorgiou et al. presented that manual, static and dynamic analysis provides better insight into the apps' state regarding privacy and security vulnerabilities. The significance of this work is more practical to show that developers improved the apps after being informed of the issues \cite{papageorgiou2018security}. Another study on both Android and iOS apps shows that the majority of apps (95.63\%) pose some potential damage because of security and privacy violations \cite{dehling2015exploring}.
 
MMH App as an alternative health intervention is a topic of great interest, but mostly it is limited to manual studies, identifying concerns, and suggesting general improvements. \bluetext{Papageorgiou et al. analyzed the security and privacy of mobile health apps covering Manual, Dynamic, Static, and Traffic analysis) \cite{papageorgiou2018security}, which is closest to our work but has different methods and analysis. For example, their work examines app permissions through manual analysis, but we performed manual and dynamic analysis to analyze app permissions. Similarly, their dynamic analysis evaluated apps based on data transmission observation over the internet. However, we analyze any exploitable app permissions, attack surface, injection, and other vulnerabilities on Android devices at runtime. Our static analysis uncovers Hardware ID Usage and other vulnerabilities different from their analysis. We also analyzed apps' privacy policies through DL-based tool \textit{Polisis} in addition to a manual analysis done in \cite{papageorgiou2018security}.} 

\label{sec:Study Design}
\section{Study Goals and Design}
We aim to evaluate Android MMH Apps to keep their security and privacy practices transparent to the users. We also look at the vulnerabilities associated with these apps, leading to security and privacy exploitation. App permissions are studied to understand the necessity of these permissions for app functionality and users' options to opt out. 
\subsection{Study Goals}
The study presents a comprehensive picture of security and privacy practices, issues, and vulnerabilities to exploit by studying privacy policies, terms and conditions, static analysis, and dynamic analysis of MMH Apps. We outline our higher-level study goals as follows:
\begin{itemize}
	\item Manual and \textit{Polisis} Analysis: Identify the deficiencies in privacy policies and challenges for users and the developers
	\item Dynamic Analysis: Identifying the threat surface and runtime vulnerabilities
	\item Static Analysis: Identifying the security and privacy threats in apps' code
	\item Traffic Analysis: Identifying the insecure data transactions
	
\end{itemize}

We comprehensively study privacy and security issues in MMH apps by combining observations from manual app usage from a regular user's perspective, privacy policy analysis, and dynamic and static analysis. We look at the exploitable app permissions levels, injection vulnerabilities, security, and privacy settings through static and dynamic analysis.

\begin{table*}[!ht]
	\centering
	\caption{Twenty Highly-Popular (HP) MMH Apps with Category, Downloads, Rating, and Rated By}
	\label{tab:hl20}
	\begin{tabular}{|L{.3cm}|L{4.5cm}|L{2.4cm}|R{1.9cm}|R{1cm}|R{1cm}|}
		\hline
		\textbf{} & \textbf{Apps}     & \textbf{Categories} & \textbf{Downloads(+)} & \textbf{Ratings} & \textbf{Rated By} \\ \hline
		1  & 7 Cups      & Depression   & 1,000,000  & 4.3  & 18,215  \\ \hline
		2  & Anxiety Relief Hypnosis    & Anxiety/ stress  & 100,000  & 4.2  & 1,375  \\ \hline
		3  & BetterHelp     & Anxiety/ stress  & 500,000  & 4.5  & 10,290  \\ \hline
		4  & Breathe2Relax     & PTSD   & 100,000  & 3.3  & 1,082  \\ \hline
		5  & Calm      & Mindfulness/ Meditation & 10,000,000  & 4.4  & 271,581  \\ \hline
		6  & CBT Thought Record Diary   & Anxiety/ stress  & 100,000  & 4.7  & 1,431  \\ \hline
		7  & eMoods      & Bipolar Disorder  & 100,000  & 4.6  & 4,133  \\ \hline
		8  & Happify      & Depression   & 1,000,000  & 4  & 2,408  \\ \hline
		9  & Headspace     & Mindfulness/ Meditation & 10,000,000  & 3.5  & 133,877  \\ \hline
		10 & MindDoc: Mood Tracker for Depression \& Anxiety & Depression   & 1,000,000  & 4.5  & 35,016  \\ \hline
		11 & MindShift     & Anxiety/ stress  & 100,000  & 4.1  & 1,227  \\ \hline
		12 & MoodSpace - Stress, anxiety   & Anxiety/ stress  & 100,000  & 4.7  & 2,915  \\ \hline
		13 & MoodTools     & Depression   & 100,000  & 4.3  & 3,108  \\ \hline
		14 & PTSD Coach     & PTSD   & 100,000  & 4.6  & 575  \\ \hline
		15 & Sanvello     & Anxiety/ stress  & 1,000,000  & 4.6  & 17,005  \\ \hline
		16 & Self-Help for Anxiety Management (SAM)  & Anxiety and stress  & 500,000  & 3.9  & 2,957  \\ \hline
		17 & Super Better     & PTSD   & 100,000  & 4.4  & 5,999  \\ \hline
		18 & Ten Percent Happier    & Mindfulness/ Meditation & 500,000  & 4.8  & 10,541  \\ \hline
		19 & What’s Up     & Suicide Prevention  & 500,000  & 4  & 3,221  \\ \hline
		20 & Wysa: stress, depression   & Anxiety/ stress  & 1,000,000  & 4.7  & 51,746  \\ \hline
	\end{tabular}
\end{table*}

\subsection{MMH App Inclusion Criteria}
As many MMH Apps are available, we study 40 MMH Apps fulfilling the following inclusion criteria. We searched online for the available mental health apps in the Android/Google Play Store. All selected MMH apps are in the English language and free to download. The primary criteria for inclusion is based on the app ratings, the number of downloads, and the number of reviewers.

\begin{table*}[!ht]
	\centering
	\vspace{0mm}
	\caption{Twenty Less-Popular (LP) MMH Apps with Downloads, Rating, and Rated By}
	\label{tab:ll20}
	\begin{tabular}{|L{.3cm}|L{4.5cm}|L{2.4cm}|R{1.9cm}|R{1cm}|R{1cm}|}
		\hline
		\textbf{\#} & \textbf{Apps}          & \textbf{Categories} & \textbf{Downloads(+)} & \textbf{Ratings} & \textbf{Rated By} \\ \hline
		1   & Anger Management \& stress relief game (pstd)  & Anxiety/ stress  & 100,000    & 2.9    & 592    \\ \hline
		2   & Bipolar Test          & Bipolar Disorder  & 10,000    & 3.1    & 41    \\ \hline
		3   & Brain Manager by UPMC        & Depression    & 1,000     & 2.1    & 18    \\ \hline
		4   & Course of Cognitive Behavioral Therapy    & Depression    & 5,000     & 3.0    & 33    \\ \hline
		5   & Daylight - Worry Less        & Anxiety/ stress  & 10,000    & 2.5    & 40    \\ \hline
		6   & Depression           & Depression    & 1,000     & 1.0    & 3     \\ \hline
		7   & Depression: The Game        & Depression    & 10,000    & 3.1    & 418    \\ \hline
		8   & EAP In Your Pocket         & Anxiety/ stress  & 5,000     & 2.3    & 16    \\ \hline
		9   & iPrevail: Anxiety \& Depression      & Anxiety/ stress  & 10,000    & 2.9    & 130    \\ \hline
		10   & Mental Health - psychologist      & Depression    & 5,000     & 3.0    & 10    \\ \hline
		11   & NarcStop - Narcissistic abuse and recovery guide & Depression    & 1,000     & 2.7    & 14    \\ \hline
		12   & PTSD Aid           & Depression    & 1,000     & 2.8    & 14    \\ \hline
		13   & R U Suicidal?          & Suicide Prevention & 1,000     & 2.6    & 33    \\ \hline
		14   & Real Antistress Stress Relief: Relaxing games  & Anxiety/ stress   & 100,000    & 3.2    & 1080    \\ \hline
		15   & SafetyNet: Your Suicide Prevention App    & Suicide Prevention & 1,000     & 2.8    & 11    \\ \hline
		16   & Stress relief ducky: antidepressant \& anti anxiety & Anxiety/ stress  & 1,000     & 3.0    & 10    \\ \hline
		17   & Suicide Prevention        & Suicide Prevention & 1,000     & 3.2    & 13    \\ \hline
		18   & Talkspace Counseling \& Therapy      & Depression    & 100,000    & 2.1    & 2774    \\ \hline
		19   & VA Health Chat          & PTSD     & 10,000    & 2.7    & 73    \\ \hline
		20   & Waver - Meet others with same Mental Health Issues & Depression    & 5,000     & 3.1    & 172    \\ \hline
	\end{tabular}
\end{table*}
We grouped apps into Highly Popular (HP) and Less Popular (LP) based on their ranking and number of downloads following inclusion criteria. Table \ref{tab:hl20} shows 20 HP apps in the range of ranking from 3.3 to 4.8, and the number of downloads varies from 100,000 to 10,000,000. For LP Apps, Table \ref{tab:ll20} shows that ratings are in the range of 1.0 to 3.2, and downloads vary from 1,000 to 100,000. Higher ratings and number of downloads were combined with a higher number of reviewers. Minimum reviewers for HP apps are 575, while the lowest LP app, with a rating of 1, has only three reviewers.

\label{sec:Data collection and analysis}
\section{Methodology and Data Collection}

We divide the study data collection into two major phases to achieve comprehensive results. In the first phase, we searched for popular and widely promoted mental health apps through different online app reviewing websites, including recommendations from creditworthy sources like the National Institute of Health (NIH) and the Americans with Disabilities Act (ADA). We categorize studied apps in 6 major categories as \textit{Suicide Prevention}, \textit{Anxiety and Stress}, \textit{Bipolar Disorder}, \textit{Depression}, \textit{PTSD}, and \textit{ Mindfulness and Meditation}.

We studied 20 highly popular (HP) mental health apps within Google Play in August 2020, and 20 less popular (LP) apps were studied in October 2020. In this study, we focus only on Android MMH apps. We shortlisted 20 apps as per inclusion criteria and collected data on the availability of the privacy policy. The privacy policies were studied manually for covered content and content inconsistencies. We downloaded all selected apps on an Android phone to observe the app’s behavior regarding the accessibility of the privacy policy. Another vital data collected is about the apps’ declaration of information being collected, data sharing, and if users can opt-out or not. We also studied the 20 LP MMH apps with lower ratings and fewer downloads to analyze the privacy policies. 

In the second phase, we focus on 20 HP apps to conduct Static, Dynamic, and Traffic analyses to identify attack surface and security and privacy vulnerabilities in highly popular MMH apps as these  may impact larger population in comparison to less popular MMH apps. We extended the Manual Analysis of privacy policies by analyzing HP apps through automated online Deep Learning (DL) framework \textit{Polisis}.

\textbf{Polisis: } Polisis\footnote{\href{https://pribot.org/polisis}{\textit{Polisis Web Interface}}.}, introduced in 2018, is a web-based tool that can analyze individual privacy policy or compare multiple policies on high-level aspects as well as fine-grained details of privacy practices. It provides a visual representation of any privacy policy for quick reference without the need of reading it manually \cite{harkous2018polisis}. 

\textbf{Static Analysis: }
Static Analysis or code analysis is examining the source code of an application to highlight possible vulnerabilities without running the code. Static Analysis inspects the code quality and helps determine multiple unknown or unseen issues that might reflect the application’s performance, availability, or vulnerabilities. We acquired the latest APK file for each app studied through an Android mobile app \textit{APK Extractor}. We used an online APK decompiler\footnote{\href{http://www.javadecompilers.com/}{\textit{APK decompiler}}.} to decompile extracted APK files. Static Analysis was performed on decompiled APK files using Android Studio 3.6.1 to review the code for white-box testing.

\textbf{Dynamic Analysis: }
Dynamic Analysis is mainly for finding run-time security vulnerabilities in a program, including identifying security flaws covering the most prevalent types of attack, such as disclosing data in transit, authentication and authorization issues, and configuration errors. We utilized a well-known Android dynamic testing tool \textit{Drozer 2.3.4} \cite{drozer}. Drozer is a highly interactive open-source security assessment tool for assessing Android apps. Drozer has a client agent installed on the test Android phone, and command-line-based tests were executed on Windows 10 machine. Drozer can identify and interact with the vulnerabilities exposed by Android apps. Developers can use Drozer to discover and minimize the potential security, and privacy risks in Android apps \cite{drozer}. 

\textbf{Traffic Analysis: } 
For Traffic Analysis, we used \textit{Android PCAP Capture} installed on the test mobile Android phone to examine the apps. We captured all the traffic for all apps by interacting with the studied MMH HP Apps when PCAP was running. We analyzed the captured PCAP files for observing any traffic vulnerabilities.

\label{sec:results}
\section{Data Analysis and Results}
We summarize our findings under manual analysis, privacy policies through web-based DL tool \textit{Polisis}, dynamic analysis, static analysis, and traffic analysis to cover privacy and security issues exhibited by studied MMH apps. We study 40 MMH apps with varying ratings and downloads classified under six categories. \textit{Anxiety and Stress} (14) and \textit{Depression} (13) are two major categories. \textit{Suicide Prevention} and \textit{PTSD} each have 4 apps, 3 \textit{Mindfulness and Meditation} apps, and 2 apps for \textit{Bipolar Disorder} are studied.

\subsection{Manual Analysis}

We document the availability and accessibility of the Privacy Policy for the studied MMH apps. We also compare our HP (20) and LP (20) app groups to see any identifiable differences. Based on ratings and the number of downloads, results show that 95\% of HP apps have a defined PP compared to 65\% of LP apps. From the privacy policies of 20 HP apps, we have the following observations: 

\subsubsection{Privacy Policy}
Sixty-five percent of 20 HP apps collect device information, while 40\% apps collect the most common data points like name, email, phone, and address. 55\% apps collect email addresses. A declaration to collect location or IP address is stated only by 15\%. 5\% apps are collecting medication information as well as counseling sessions. 

On personal data security and privacy, only 30\% apps maintain de-identified data, 10\% apps declare not de-identifying data, and 60\% apps do not provide information on de-identifying data. Most HP apps share data with third-party providers, and 20\% declare to share data with sales and legal teams. only 10\% apps share data with therapists or healthcare providers. 60\% apps store data with the app provider, 30\% store on the user’s device, and 5\% apps store data on the cloud. 50\% apps allow users to opt-out, 15\% apps do not allow it, and 35\% do not provide any information on the data sharing opt-out option. 40\% apps offer security through account ID and password, while 20\% provide encryption and firewalls. However, few apps mention providing secure access with the paid version of the app. One app mentions \textit{commercial means security}, but no further detail is provided.

Few app-specific privacy policies indicate threats to the privacy and security of the app user. For example, \textit{7 Cups} says, “While we generally do not monitor transcripts of chats between users and Listeners and Therapists, we may occasionally review the chat transcripts to conduct quality control, address potential safety issues, and prevent misuse of our platform, if certain suspicious or potentially harmful activity is detected.” App also says that the ‘Do Not Track’ feature is not supported, so users cannot opt out. App \textit{Calm} also suggests that the app will retain some information as required or permitted by law for a certain period even after the user requests to cancel or delete the account.
 
\begin{figure*}[t]
	\centering
	\vspace{-5mm}
	\includegraphics[width=.75\linewidth]{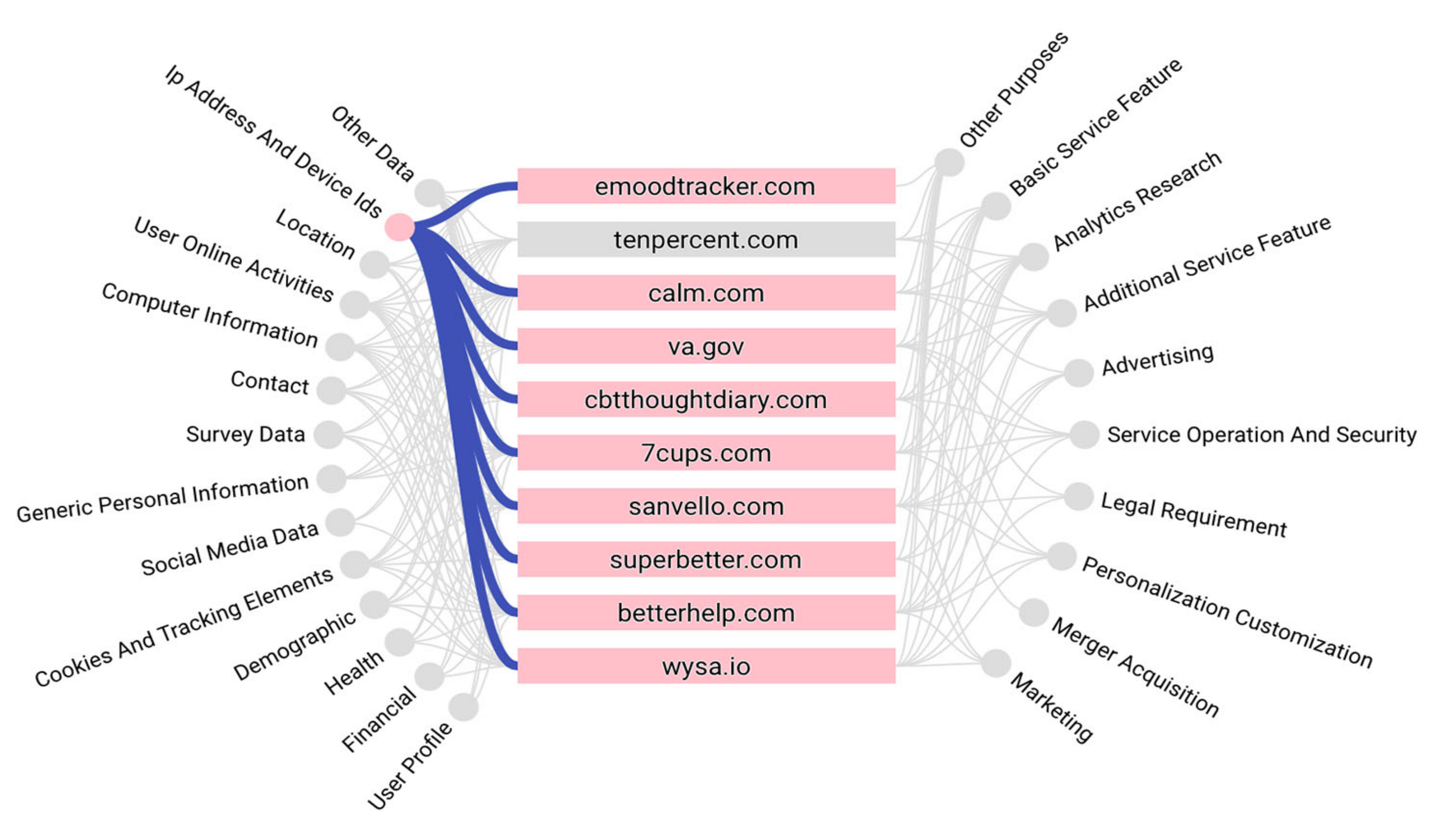}
	\vspace{0 mm}
	\caption{Polisis \cite{harkous2018polisis} Analysis- Comparison of HP MMH Apps Privacy Policies}
	\label{fig:polisistenapps}
\end{figure*}
\subsection{Privacy Policy Analysis using Polisis}
Polisis allows up to 10 privacy policies to compare, and for 10 HP MMH apps' privacy policies, we found that 65\% of the applications do not discuss Secure Data storage, Privacy Security Program, and Security Data program. 60\% of the applications have not mentioned data access limitations and secure user authentication. Also, 40\% of the applications are unclear about their security measure, and 20\% of them use many generic statements, which are confusing for the readers. Fig. \ref{fig:polisistenapps} demonstrates the Polisis comparison of 10 MMH apps’ privacy policies on key data collection (Left side) and data usage (Right side) categories. For example, 9 out of 10 HP MMH apps are collecting Device IDs and IP Addresses which can be exploited to violate users’ privacy and security. Only 5 out of 10 apps use collected information for service operations and security. All of the apps have a date category as \textit{Other Data}, and all the apps may use the collected information for \textit{Other Purposes}, which hinders the transparency about what is being collected and what it can be used for. 
\begin{figure*}[htb]
	\centering
	\vspace{2mm}
	\includegraphics[width=.70\linewidth]{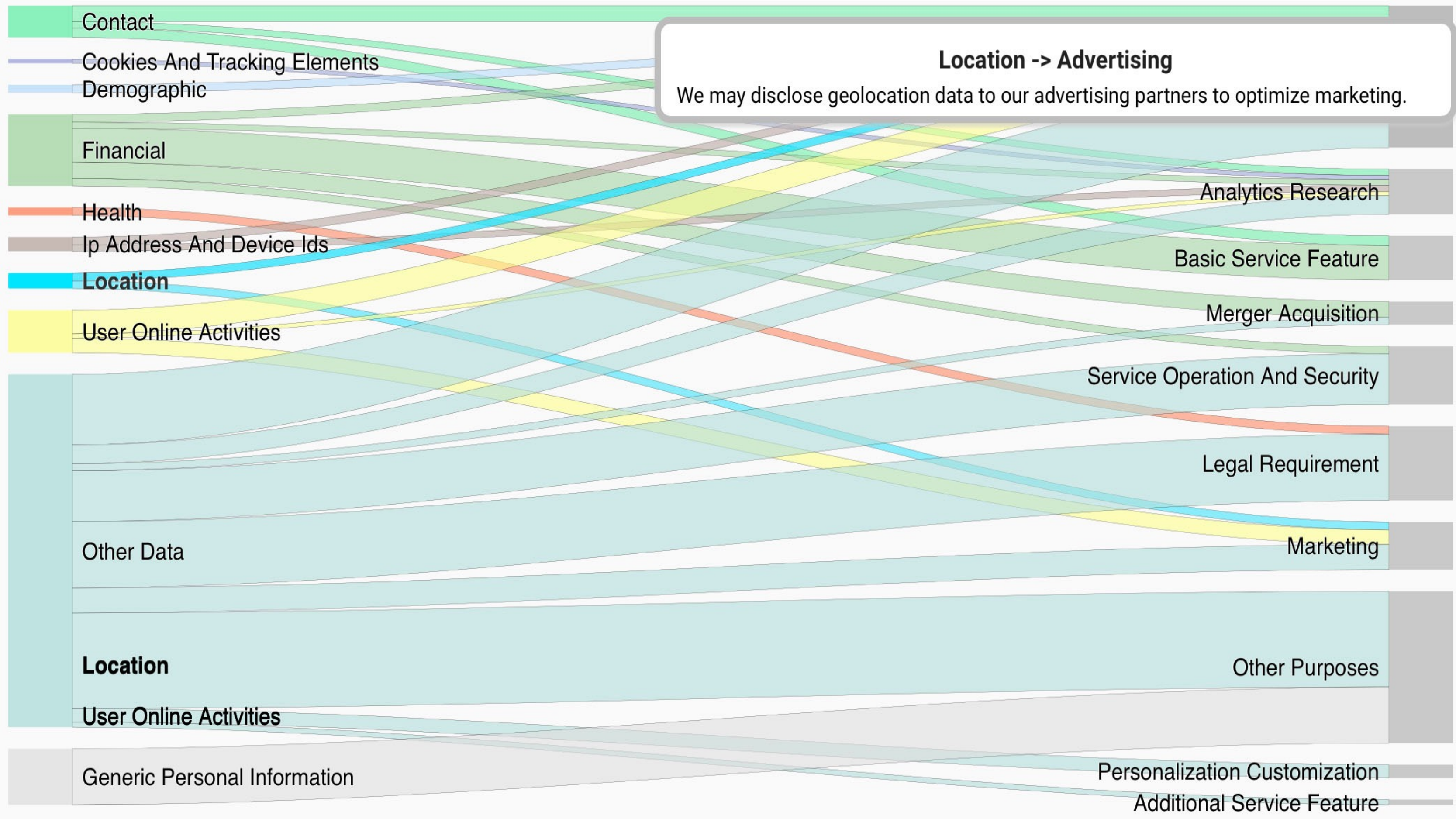}
	\caption{Polisis Analysis of MMH App \textit{BetterHelp}- Data Collection and Data Usage}
	\label{fig:polisis}
\end{figure*}
Fig. \ref{fig:polisis} represents the details of a specific app (\textit{Betterhelp}) regarding information collection (left side) and how the collected information can be utilized (right side). We can see that `Other Data’ goes to many different usages, including `Other Purpose’, making it difficult for users to understand the associated risks. We observe that 9 out of 10 compared apps pay attention to children’s privacy as they represent a more vulnerable population. However, 8 out of 10 do not talk about sharing personal information with third parties, and two apps show concern with \bluetext{warning signs stating that “Several types of personal information types are shared with third parties.”}

\begin{table}[!thb]
	\centering
	
	\caption{Static Analysis: Major Security and Privacy Vulnerabilities in MMH Apps}
	\vspace{1mm}
	\label{tab:staticsecurity}
	\begin{tabular}{|L{2.9cm}|L{1cm}|L{.9cm}|L{1.1cm}|L{.9cm}|}
		\hline
		\textbf{App}    & \textbf{\begin{tabular}[c]{@{}c@{}}Hardware\\ ID\\ Usage\end{tabular}} & \textbf{\begin{tabular}[c]{@{}c@{}}Insecure \\ TLS/SSL\end{tabular}} & \textbf{\begin{tabular}[c]{@{}c@{}}Potential\\ Multiple \\ Certificate\\ Exploit\end{tabular}} & \textbf{\begin{tabular}[c]{@{}c@{}}File \\ Readable\end{tabular}} \\ \hline
		7 Cups     & \cmark         & \cmark         & \cmark            &\cmark         \\ \hline
		Anxiety Relief Hypnosis   & \xmark         & \xmark         & \xmark            & \cmark         \\ \hline
		BetterHelp    & \xmark         & \xmark         & \xmark            &\xmark         \\ \hline
		Breathe2Relax    & \cmark         & \xmark         & \cmark            &\xmark         \\ \hline
		Calm     & \xmark         & \cmark         & \xmark            &\xmark         \\ \hline
		CBT Thought Record Diary   & \cmark         & \xmark         & \cmark            &\xmark         \\ \hline
		eMoods     & \cmark         & \xmark         & \cmark            &\cmark         \\ \hline
		Happify     & \cmark         & \cmark         & \cmark            &\cmark         \\ \hline
		Headspace    & \cmark         & \xmark         & \cmark            &\cmark         \\ \hline
		MindDoc     & \xmark         & \xmark         & \xmark            &\xmark         \\ \hline
		MindShift    & \xmark         & \xmark         & \xmark            &\xmark         \\ \hline
		MoodSpace - Stress, anxiety  & \xmark         & \xmark         & \xmark            &\xmark         \\ \hline
		MoodTools    & \cmark         & \xmark         & \cmark            &\xmark         \\ \hline
		PTSD Coach    & \xmark         & \xmark         & \xmark            &\xmark         \\ \hline
		Sanvello     & \xmark         & \xmark         & \xmark            &\xmark         \\ \hline
		Self-Help for Anxiety Management (SAM) & \cmark         & \xmark         & \cmark            &\xmark         \\ \hline
		Super Better    & \cmark         & \cmark         & \cmark            &\cmark         \\ \hline
		Ten Percent Happier   & \xmark         & \xmark         & \xmark            & \cmark         \\ \hline
		What’s Up    & \xmark         & \xmark         & \xmark            &\xmark         \\ \hline
		Wysa: stress, depression   & \xmark         & \xmark         & \cmark            &\xmark         \\ \hline
	\end{tabular}
\end{table}

\subsection{Static Analysis}
 We performed a Static Analysis on HP MMH apps studied in our work. In addition, we study the security issues and vulnerabilities existing in code analysis that may be exploited in security and privacy attacks.
 
Broadly, as categorized in Android Studio 3.6.1, we observed errors and warnings for five categories of \textit{Accessibility}, \textit{Correctness}, \textit{Performance}, \textit{Usability}, and \textit{Security}. We further focused on identifying major security flaws. Table \ref{tab:staticsecurity} shows four studied categories of security and privacy features that can potentially be exploited. Thirteen out of twenty HP MMH apps are vulnerable to at least one category, and three apps show vulnerability to all four categories. Our results show that 45\% of HP MMH apps show \textit{Hardware Id Usage}, a unique permanent ID (Android hardware ID can get reset with a factory reset option) for the device and thus can be associated with a particular user. User Hardware ID can be accessible to all apps installed on the device as per the Android framework scope, which increases the risk of privacy violation \cite{AndroidDeveloper}.

50\% apps indicate \textit{Potential multiple certificate exploits}, showing that App signatures can be exploited if not validated properly. 30\% apps can read the files through \textit{File.setReadable()}, which is used to make file word-readable. 20\% apps show \textit{Insecure TLS/SSL Trust Manager}, which can cause insecure network traffic disclosing sensitive information. TrustManager can implement custom certificate validation strategies, and an insecure trust manager implementation makes an application vulnerable to Man-In-The-Middle attacks. Malicious entities may intercept an app’s data over the network with insecure TLS/SSL and can violate the users’ privacy and security \cite{AndroidDeveloper}.

\begin{figure*}[!htb]
	\centering
	\includegraphics[width=.85\textwidth]{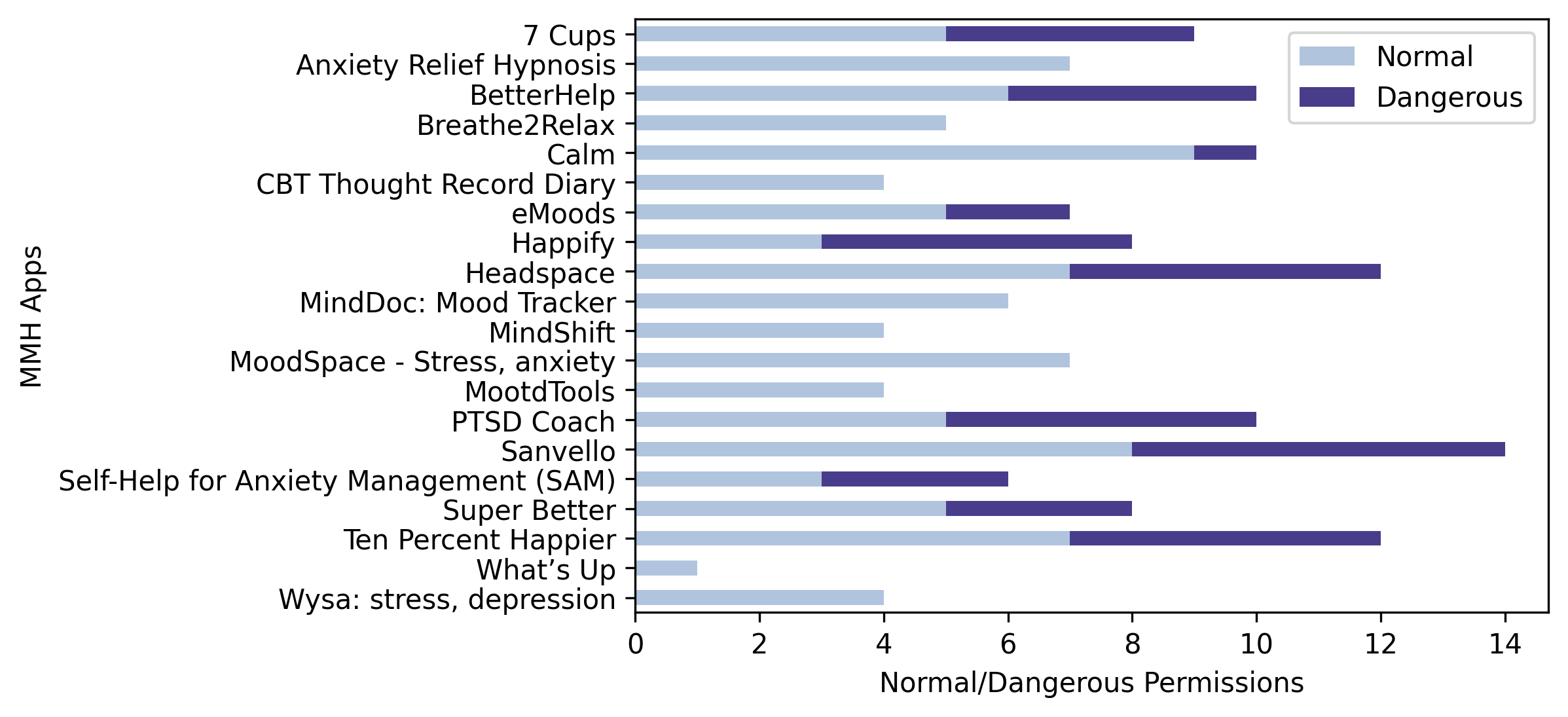}
	\caption{Dynamic Analysis- Permissions Used by HP MMH Apps}
	\label{fig:dapermissions}
\end{figure*}

\subsection{Dynamic Analysis}
Dynamic analysis is conducted on twenty HP apps to analyze the app permissions, attack surface, and exploitable vulnerabilities on Android devices at runtime.
\subsubsection{Permissions}
App permissions analysis can identify the potential risk of violating users’ privacy and security. Standard permissions allow access to data and actions which present minimal risk to the user’s privacy. Runtime permissions allow additional access to restricted data and may significantly perform restricted actions that impact the system and other apps. For example, many runtime permissions access private user data, potentially sensitive information like user’s location and contact information.

Drozer identified that more than 50\% HP apps have 1 to 6 permissions per app defined as \textit{Dangerous} according to Android developer guidelines \cite{AndroidDeveloper}. There are 12 different types of dangerous permissions identified in studied apps. From 148 standard and dangerous permissions, 29.05\% permissions fall under the dangerous category. For example, permission to read contacts threatens privacy and security as a malicious app can use this data without the user’s knowledge. Based on the permissions count of all twenty HP apps permissions, 18.38\% are listed as dangerous, while 44.87\% fall under standard permissions. The rest of the permissions consist of obsolete and other permissions. Fig. \ref{fig:dapermissions} shows the distribution of app permissions, while Table \ref{tab:dangerper} shows the specific dangerous permissions used by studies apps. \bluetext{ These highly popular MMH Apps do carry dangerous permissions, which can be exploited to evade patients’ privacy. Therefore, it is critical to define who may access such data and its maintenance.}

\begin{table}[htb]
	\centering
	\vspace{0mm}
	\caption{Dynamic Analysis: Dangerous Permissions}
	\vspace{1mm}
	\label{tab:dangerper}
	\begin{tabular}{|L{6cm}|R{1cm}|}
		\hline
		\textbf{App Permission (Category- Dangerous)} & \textbf{HP Apps}\\ \hline
		android.permission.WRITE\_EXTERNAL\_STORAGE & 9  \\ \hline
		android.permission.READ\_CONTACTS  & 3  \\ \hline
		android.permission.READ\_EXTERNAL\_STORAGE & 9  \\ \hline
		android.permissions.RECORD\_AUDIO  & 4  \\ \hline
		android.permission.READ\_CALENDAR  & 1  \\ \hline
		android.permission.WRITE\_CALENDAR  & 1  \\ \hline
		android.permission.READ\_PHONE\_STATE  & 5  \\ \hline
		android.permission.GET\_ACCOUNTS  & 2  \\ \hline
		android.permission.CAMERA   & 4  \\ \hline
		android.permission.ACCESS\_FINE\_LOCATION & 2  \\ \hline
		android.permission.ACCESS\_COARSE\_LOCATION & 2  \\ \hline
		android.permission.BODY\_SENSORS  & 1  \\ \hline
		
	\end{tabular}
\end{table}
We also manually observed all apps installed on Android phone and requested permissions. Table \ref{tab:dphp20} presents eight major permissions. Results show that 12 out of 20 HP MMH apps carry 31 dangerous permissions. Storage Access permission is most common, with ten apps requesting it, while body sensors and calendar access are the least requested (1 app each) permission. Common requested permissions are for Camera(4), Microphone (5), and Contacts (5). Other requested permissions are Phone (3) and Location (2). These findings are per dynamic analysis identified dangerous permissions. There are two differences observed between manual and dynamic analysis. First is that app \textit{BetterHelp} does not show read or write storage permissions in Drozer output, but it is requesting `Storage’ permission on an Android phone. Second is app \textit{Wysa} which doe not show `RECORD\_AUDIO’ permission in Drozer output, but on the phone, `Microphone’ permission is requested.

\begin{table*}[]
\centering
\vspace{0mm}
\caption{Manual Analysis: Dangerous Permissions Requested for 20 HP MMH Apps }
\vspace{1mm}
	\label{tab:dphp20}
\begin{tabular}{|L{3cm}|P{1cm}|P{1.2cm}|P{1.2cm}|P{.5cm}|P{1.2cm}|P{1.1cm}|P{1cm}|P{1cm}|}
\hline
\textbf{\begin{tabular}[c]{@{}l@{}}App\end{tabular}} & \textbf{Camera} & \textbf{Location} & \textbf{\begin{tabular}[c]{@{}l@{}}Body \\ sensors\end{tabular}} & \textbf{Mic} & \textbf{Contacts} & \textbf{Calendar} & \textbf{Phone} & \textbf{Storage} \\ \hline
7 Cups                   & \xmark     & \xmark     & \xmark                 & \xmark      & \xmark     & \xmark     &\xmark    & \cmark                         \\ \hline
Anxiety Relief Hypnosis               &\xmark     & \xmark     &\xmark                 &\xmark      &\xmark     &\xmark     &\xmark    &\xmark                         \\ \hline
BetterHelp                  & \cmark    &\xmark     &\xmark                 & \cmark     & \cmark     &\xmark     &\xmark    & \cmark                         \\ \hline
Breathe2Relax                 &\xmark     &\xmark     &\xmark                 &\xmark      &\xmark     &\xmark     &\xmark    &\xmark                         \\ \hline
Calm                   &\xmark     &\xmark     &\xmark                 &\xmark      &\xmark     &\xmark     & \cmark    & \xmark                         \\ \hline
CBT Thought Diary              &\xmark     &\xmark     &\xmark                 & \xmark     &\xmark     &\xmark     &\xmark    & \xmark                        \\ \hline
eMoods                   &\xmark     &\xmark     &\xmark                 &\xmark      &\xmark     &\xmark     &\xmark    & \cmark                         \\ \hline
Happify                   & \cmark    & \cmark     & \cmark                &\xmark      & \cmark     &\xmark     &\xmark    & \cmark                         \\ \hline
Headspace                  &\xmark     &\xmark     & \xmark                 &\xmark      &\xmark     & \cmark     & \cmark    & \cmark                         \\ \hline
MindDoc                   &\xmark     &\xmark     &\xmark                 &\xmark      &\xmark     &\xmark     &\xmark    & \xmark                        \\ \hline
MindShift                  &\xmark     &\xmark     & \xmark                 &\xmark      &\xmark     &     \xmark&    &\xmark                         \\ \hline
MoodSpace              &\xmark     &\xmark     &\xmark                 &\xmark      &\xmark     &\xmark     &\xmark    &\xmark                         \\ \hline
MootdTools                  &\xmark     &\xmark     &\xmark                 &\xmark      &\xmark     &\xmark     &\xmark    &\xmark                         \\ \hline
PTSD Coach                  & \cmark    &\xmark     &\xmark                 & \cmark     & \cmark     &\xmark     &\xmark    & \cmark                         \\ \hline
Sam                    &\xmark     &\xmark     &\xmark                 & \cmark     &\xmark     &\xmark     &\xmark    & \cmark                         \\ \hline
Sanvello                  & \cmark    &\xmark     &\xmark                 & \cmark     & \cmark     &\xmark     &\xmark    & \cmark                         \\ \hline
SuperBetter                  &\xmark     &\xmark     &\xmark                 &\xmark      & \cmark     &\xmark     &\xmark    & \cmark                         \\ \hline
Ten percent happier                &\xmark     & \cmark     & \xmark                 &\xmark      & \xmark     &\xmark     & \cmark    & \cmark                         \\ \hline
What’s up                  &\xmark     & \xmark     &\xmark                 &\xmark      &\xmark     &\xmark     &\xmark    &\xmark                         \\ \hline
Wysa                   &\xmark     &\xmark     &\xmark                 & \cmark     &\xmark     &\xmark     &\xmark    &\xmark                          \\ \hline
Total=                   &4     &2     & 1                 & 5     &5     &1     &3    &10     
\\ \hline  
\end{tabular}

\end{table*}

\begin{figure*}[]
	\centering
	\vspace{2mm}
	\includegraphics[width=.9\linewidth]{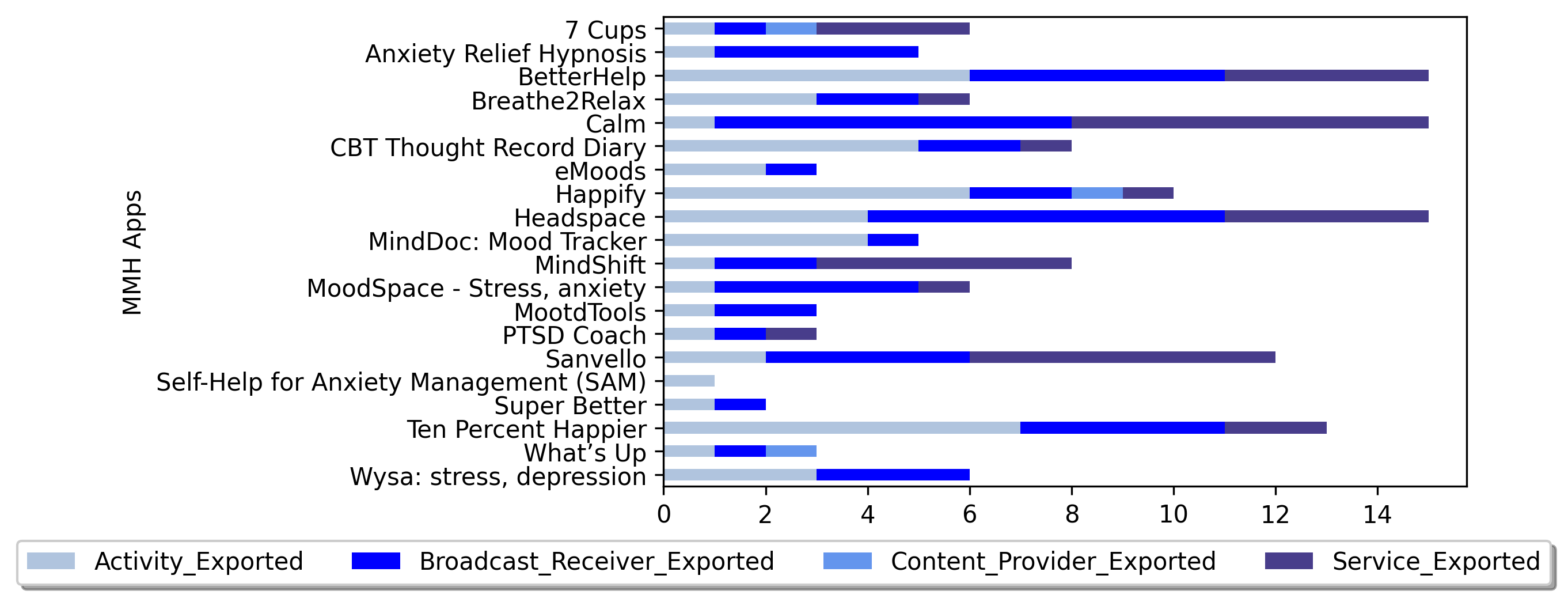}
	\caption{Dynamic Analysis - HP MMH Apps Attack Surface}
	\label{fig:attacksurface}
\end{figure*}

\subsubsection{Attack Surface}
As per Drozer output, all HP MMH apps show exposure as an attack surface. Attack surface is categorized as `Activities Exported’, `Broadcast Receiver Exported’, `Content Providers Exported’, and `Service Exported’. 

Fig. \ref{fig:attacksurface} shows how each HP app has entry/exit points as attack surfaces under these four categories that can be exploited by a malicious user or a malicious app. There are a total of 52 activities exported for 20 studied apps. Another query in Drozer indicates if any permission is required or not-required to perform some app activity. If no permission is required to export some activity, any application will launch the activity, allowing a malicious application to gain access to sensitive information. It also allows to modify the internal state of the application or make the user interact with the victim application while believing that the user is still interacting with the malicious application \cite{cwe}. 

As per our results, there are 36 services exported, and three content providers are exported. If access to an exported Service is not restricted, any application may start and bind to the Service. The exposed functionality may allow a malicious application to perform unauthorized actions like gaining access to sensitive information or corrupting the application’s internal state. If access to a Content Provider is not restricted to only the expected applications, then malicious applications might be able to access the sensitive data \cite{cwe}.
\begin{figure}[]
	\centering
	\vspace{0mm}
	\includegraphics[width=.9\linewidth, height=7cm]{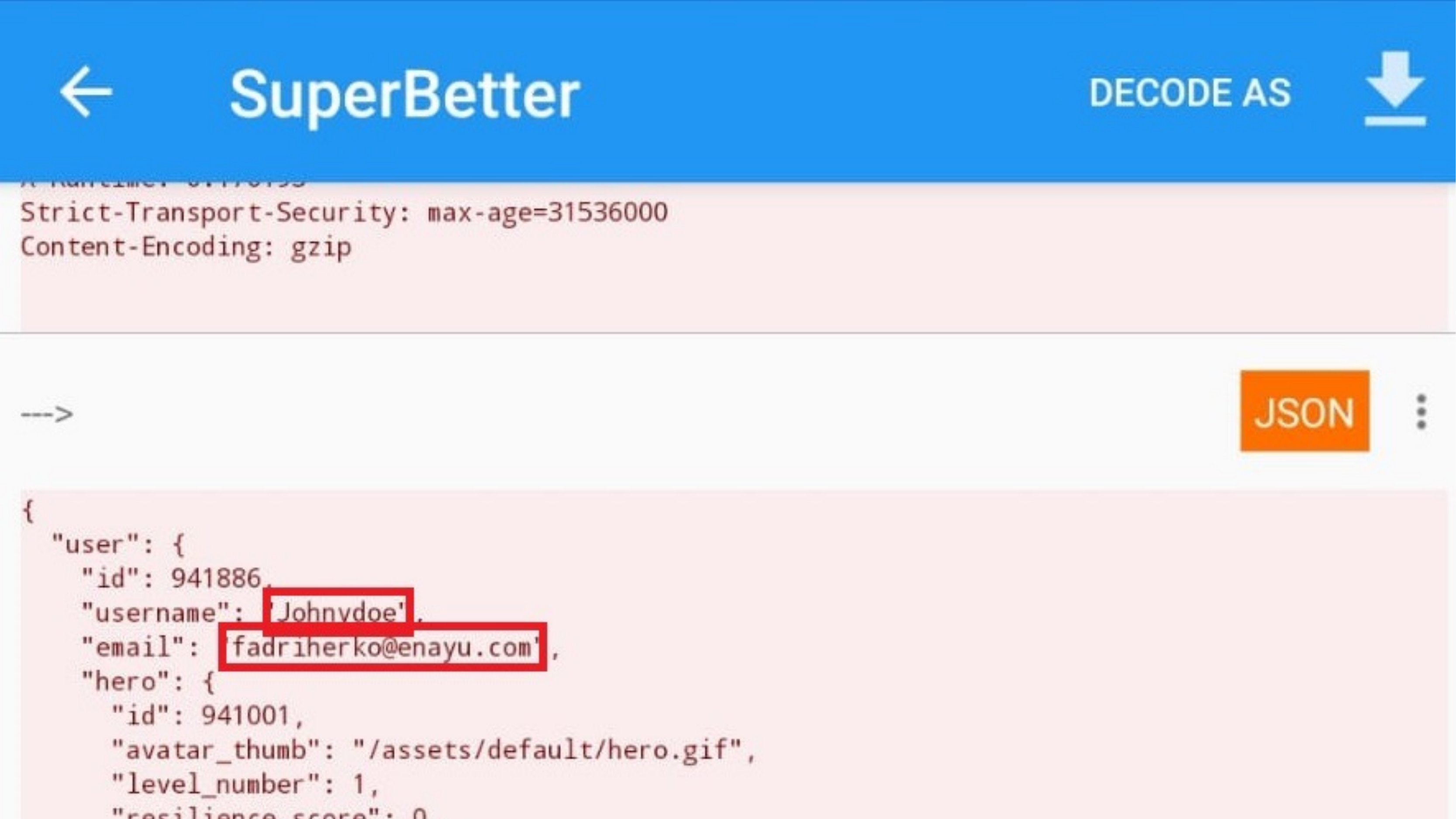}
	\vspace{0mm}
	\caption{\textit{SuperBetter}: Traffic captured through `pcap' shows username and email in plaintext}
	\label{fig:sbtext}
\end{figure}

\begin{figure}[]
	\centering
	\vspace{0mm}
	\includegraphics[width=.9\linewidth]{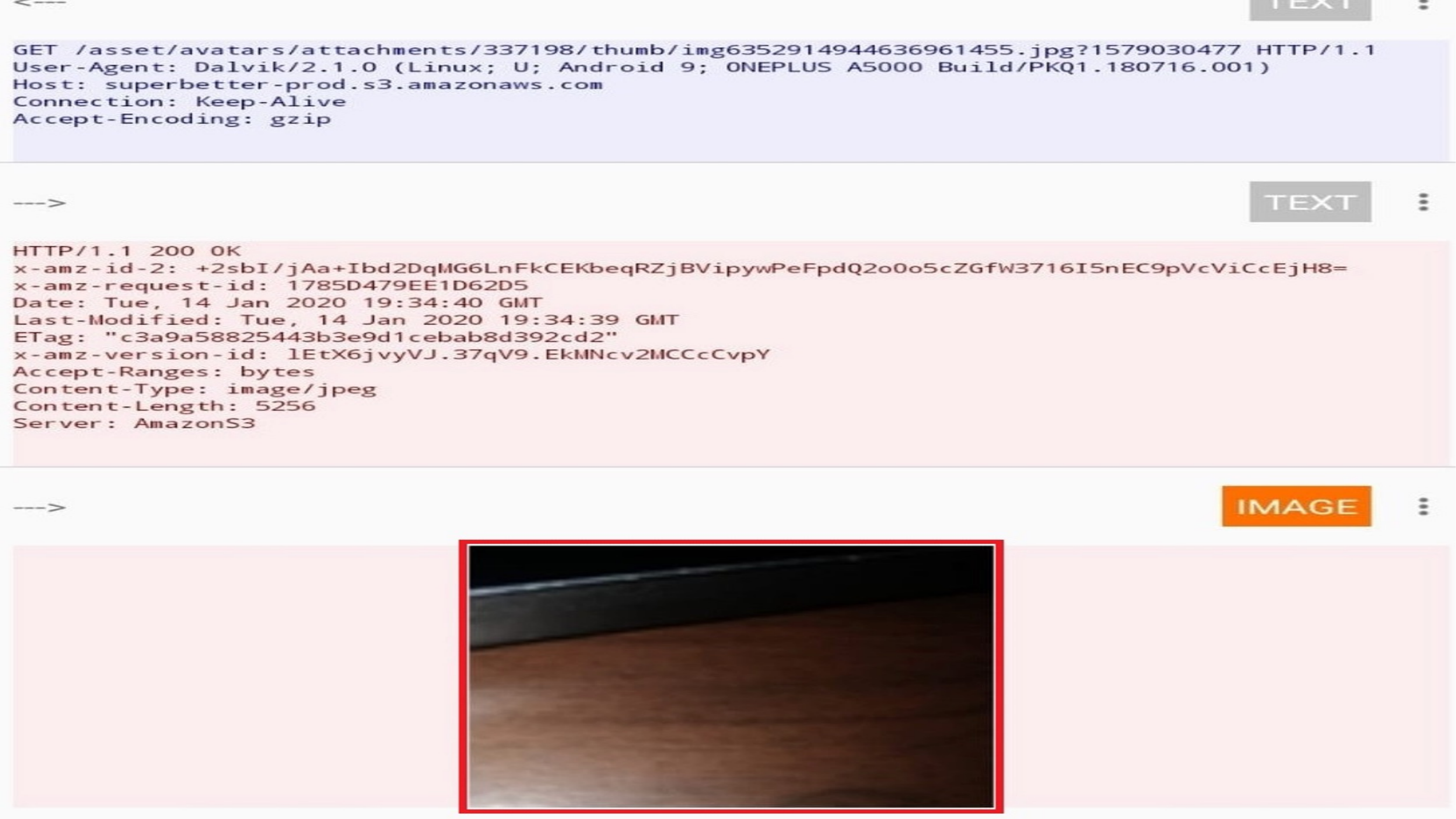}
	\caption{\textit{SuperBetter}: Traffic captured through pcap shows Unencrypted Image }
	\label{fig:sbimage}
\end{figure}
Our findings also show 54 \textit{Broadcast Receiver Exported}, which is another exploitable vulnerability. Additionally, a malicious application can send an explicit intent to the victim app, which may assume that any received intent is valid, which may cause unintended app behavior.

\subsubsection{Injection and Other Vulnerabilities}
We scanned all 20 HP apps for injection vulnerabilities, but Drozer did not identify any app as vulnerable. However, if the content provider is exported and \textit{grantUriPermissions} is set true, that can make an app vulnerable to injection attacks. Drozer indicated three such apps where we have null permissions and content provider is exported, but all three apps have \textit{grantUriPermissions} as false. We scanned 20 HP MMH apps to find any vulnerable provider, but Drozer did not identify any vulnerable provider for any app. Furthermore, Drozer did not identify any vulnerable URI paths for the 20 HP MMH apps.
\subsection{Traffic Analysis}
We ran each application separately and used the packet capturing tool \textit{pcap} and monitored the packets. However, currently, we could only find defects in one application, i.e., \textit{Superbetter}. Superbetter is an application for teens dealing with stress, anxiety, depression, or other mental challenges. Though we could find only one app to show this behavior, it does reflect the discrepancies between the app’s privacy policy and app behavior. As per the privacy policy of Superbetter, data is stored and encrypted, but we could capture the plain text in the pcap file as shown in Fig. \ref{fig:sbtext}. Fig. \ref{fig:sbimage} shows the unencrypted image in the captured traffic. Our experiments show that an MMH app transmits text data and images over the network without any encryption. However, information transferred over the network unprotected can be intercepted and threaten users’ privacy and security. 

\subsection{Summary of Results}
We can summarize our results as follows:
\begin{itemize}
	\item Manual Analysis: 20 HP MMH apps show that 95\% of HP apps provide a privacy policy compared to 65\% of LP apps. 65\% of apps collect device information, while 40\% collect the most common data points like name, email, phone, and address. 5\% of apps are collecting medication information and counseling sessions. Polisis analysis demonstrates that apps collect identifiable information, share it with third parties, and retain some data indefinitely. 12 out of 20 HP MMH request runtime ‘dangerous’ permissions on Android phones. 	
	\item Dynamic Analysis: More than 50\% of 20 HP MMH apps use 12 dangerous permissions. Attackers and malicious apps may exploit identified 145 vulnerabilities to access sensitive information. 
	\item Static Analysis: 45\% of MMH apps use a unique identifier, \textit{Hardware Id}, which can connect particular users to their mental health issues, thus violating privacy and security. 20\% of apps indicate \textit{Insecure TLS/SSL Trust Manager}, which may leak sensitive data through insecure network traffic. 
	\item Traffic Analysis: Captured traffic unveils the unencrypted text and image in one app while the app’s privacy policy indicates data encryption.
	
\end{itemize}

\label{sec:Limitations and Challenges}

\section{Challenges and Limitations}
In the era of ever-changing information at a fast pace, MMH apps are also subject to rapid changes in usage, ratings, downloads, and reviews from the user population. Reading privacy policies is a time-consuming process but more difficult to analyze and compare without a standard format followed by the app developers. Though tools like \textit{Polisis} can help to extract usable information automatically from the Privacy Policy, inconsistent content covered in privacy policies may limit the comparison analysis. This study is also limited to the Android platform only. \bluetext{Frequent changes in app usage and availability forced us to recollect some data.}

It is also worth mentioning that we were limited to observing the complete app functionality in unpaid versions. A complete analysis of all the app's features, especially regarding audio, video, call, and message capabilities, can provide more details on how some app utilizes or exploits app capabilities and permissions.

\label{sec:discussion and future work}

\section{Discussion and Future Work}

It is a general assumption that no user reads privacy policies because these can be out of the comprehension of an average user. Users can be more open to adopting MMH apps with more confidence if they understand the privacy policies with associated options and risks involved. \bluetext{Most MMH apps collect personal identifiable and sensitive mental health information presenting higher risks. Insecure MMH apps impact millions of users, hindering their adoption as an alternative or complementary intervention \cite{olff2015mobile}. Especially with COVID-19, mental health needs greater attention \cite{winkler2020increase}. Secure MMH apps can increase the user confidence to seek medical attention timely \cite{luxton2011mhealth}, which can help underserved populations with limited access to clinical/ non-clinical practitioners \cite{thomas2009county}. Our work stresses an immediate need to address the lack of transparent policies and development standards for MMH apps \cite{chan2015towards}.}

As a defense, developers may minimize declaring unnecessary permissions. For example, if an app needs to capture a picture using a pre-installed system camera app, instead of declaring the CAMERA permission, invoke the ACTION\_IMAGE\_CAPTURE intent action \cite{AndroidDeveloper}. To mitigate the privacy concern associated with a unique hardware id, the use of an alternate identifier, accessible by only a single app instance, can avoid the cross-app device tracking \cite{AndroidDeveloper}.  

Future work may include MMH apps on the iOS platform, identifying any significant variation in developer practices and privacy policies over time. Studying more MMH apps can evaluate ongoing efforts and take necessary steps towards further improving security and privacy. \bluetext{A user-centered study can identify practical concerns and difficulties hindering the decision to use MMH apps and may add vital inputs regarding users' preferences for adapting MMH apps as alternative interventions. }

\label{sec:conclusion}
\section{Conclusion}

Studied MMH Policies do not indicate app permissions or the necessity of permissions for intended app functionalities. Not all policies mention collecting location data, but it can be collected indirectly from the device information, cookies, and web beacons. Data sharing with many other parties is inevitable as different parties maintain the app and store data in the cloud. As a result, the app can get geolocation that is not transparent to the user, posing a privacy violation. Users may have no control over some in-app settings to opt out from specific data sharing as suggested in their privacy policies. Users may want to discontinue using the app and delete the account(s), but data can be retained for a more extended period as covered by different clauses. 

Results show that manual, static, dynamic, and traffic analysis identify exploitable vulnerabilities in MMH apps, posing privacy and security threats. More transparent privacy policies and standard secure development guidelines are needed to ensure that users and providers trust the apps. Otherwise, MMH apps can be intrusive to users’ privacy and security, negatively affecting their intended purpose of providing alternative interventions to improve their mental health.

\bibliographystyle{IEEEtran}
\bibliography{mainref}

\newpage

\end{document}